\documentclass[aps,pre,amsfonts,floatfix, twocolumn]{revtex4}%

\usepackage{graphics}
\usepackage{placeins}
\usepackage{graphicx}
\usepackage{amssymb}
\usepackage{amsmath}
\usepackage[active]{srcltx}

\date{\today}

\newcommand{\ie}{i.~e.\ }
\newcommand{\eg}{e.~g.\,\,}

\newcommand{\e}{\mathrm e}

\newcommand{\eq}[1]{(\ref{eq:#1})}
\newcommand{\fig}[1]{FIG.~\ref{fig:#1}}
\newcommand{\refS}[1]{Section \ref{sec:#1}}

\newcommand{\comment}[1]{}

\begin{document}

\title{\large Deterministic evolutionary game dynamics in finite populations}
\author{Philipp M.~Altrock and Arne Traulsen}
\affiliation{Max-Planck-Institute for Evolutionary Biology, 24306 Pl{\"o}n, Germany\\altrock@evolbio.mpg.de}
\date{\today}

\begin{abstract}

Evolutionary game dynamics describes the spreading of successful strategies in a population of reproducing individuals. 
Typically, the microscopic definition of strategy spreading is stochastic, such that the dynamics
becomes deterministic only in infinitely large populations. 
Here, we introduce a new microscopic birth--death process that has a fully deterministic strong selection limit in well--mixed populations of any size. 
Additionally, under weak selection, from this new process the frequency dependent Moran process is recovered. 
This makes it a natural extension of the usual evolutionary dynamics under weak selection. 
We analytically find simple expressions for the fixation probabilities and average fixation times of the new process in evolutionary games with two players and two strategies.  
For cyclic games with two players and three strategies, we show that the resulting deterministic dynamics crucially depends on the initial condition in a non--trivial way. 
\end{abstract}

\maketitle

\section{Introduction}\label{sec:Intro}

Evolutionary game dynamics results from the transfer of economic ideas to biology \cite{maynard-smith:1973to,maynard-smith:1982to,nowak:2004aa,nowak:2006bo}. 
In economics, rational players try to find the best strategy to maximize their payoffs. 
In biology, those individuals who use the best strategy obtain the highest reproductive fitness and spread in the population. 

Traditionally, evolutionary game dynamics is considered in infinitely large, well--mixed populations. 
This typically leads to the replicator dynamics, a system of nonlinear differential equations governing the evolutionary dynamics \citep{taylor:1978wv,zeeman:1980ze,weibull:1995hp,hofbauer:1998mm}. 
For any composition of the population, the replicator dynamics determines deterministically the direction and velocity of evolutionary dynamics. 
The replicator dynamics can be derived from microscopic models of strategy spreading, which are typically 
stochastic \cite{helbing:1993aa,helbing:1996aa,schlag:1998aa,traulsen:2005hp,traulsen:2006ab}.  
The precise definition of strategy spreading between individuals can have decisive consequences for the dynamics, in particular in structured populations \cite{szabo:2007aa,abramson:2001nx,vainstein:2001hp,ebel:2002aa,santos:2005pm,pacheco:2006pb,ohtsuki:2007pr,poncela:2008aa}.
  
Since microscopic models of strategy spreading are typically stochastic, 
evolutionary game dynamics in finite populations can only be characterized in a probabilistic way.
The most important quantities are the probability that a mutant takes over a population and the average time for this process \cite{nowak:2004pw,taylor:2004wv,antal:2006aa,taylor:2006jt}. 
Different models for strategy spreading have been proposed. 
A popular model is to choose two players, Harry and Sally, at random and to let Harry adopt the strategy of Sally with a probability given by the Fermi function, $(1+\exp[-\beta (\pi^\text{H} - \pi^\text{S})])^{-1}$, where $\pi^\text{H}$ is the payoff of Harry and $\pi^\text{S}$ is the payoff of Sally \cite{blume:1993jf,szabo:1998wv,traulsen:2006bb,traulsen:2007cc}. 
The parameter $\beta$ measures the intensity of selection. 
For $\beta \ll 1$, selection is weak and strategy spreading is essentially random. 
For $\beta \gg 1$, selection is strong and only strategies that are more successful will be 
imitated. 
For $\beta \to \infty$, the direction of the process for two strategies becomes deterministic and thus 
the fixation probability is either $0$ or $1$. 
However, even in this case, the process is only semi-deterministic, as the time of fixation remains stochastic \cite{traulsen:2007cc}.

Here, we introduce a variant of the Moran process which leads to a fully deterministic evolutionary process in finite populations under strong selection. 
For weak selection, we essentially recover the transition probabilities of the standard frequency dependent Moran process under weak selection.

We describe evolutionary game dynamics in symmetric $2 \times2$ games defined by the general payoff matrix
\begin{align}\label{eq:Pmatrix}
\bordermatrix{
  & A & B \cr
A & a & b \cr
B & c & d \cr}.
\end{align}
An $A$ player will obtain $a$ when playing against another $A$ or $b$ when playing against $B$. 
Choosing strategy $B$ results in either obtaining $c$ (against $A$) or $d$ (against $B$).

The average payoffs are obtained from pairwise interactions with all other individuals in the population of size $N$. 
This is the standard assumption and refers to the fact that the population is well--mixed, \ie there is no explicit population structure. 
Excluding self interactions, this leads to
\begin{align}
\label{eq:payoffA}
\pi^A_i &= \frac{i-1}{N-1} \,a + \frac{N-i}{N-1}\,b 
\\ 
\label{eq:payoffB}
\pi^B_i &= \frac{i}{N-1} \,c + \frac{N-i-1}{N-1} \,d,
\end{align}
where $i$ is the current number of $A$ players in the population.
Individuals with higher average payoffs produce offspring (or are imitated) with a higher probability. 
Thus, reproductive success is based on the payoff from the game.
The intensity of selection $\beta$ controls the importance of success in the game for reproductive success. 
The larger the intensity of selection, the stronger the influence of the average payoff difference on reproductive fitness.

The paper is organized in the following way. 
In \refS{EvDyn} we introduce the birth--death process as a general framework of evolutionary dynamics between two types in finite, well--mixed populations. 
In particular, we address the probability and times of absorption. 
In \refS{Moran} we give three explicit analytical forms for the microscopic dynamics and we discuss the possibility to analyze strong selection in each case. 
We show that the standard Moran process and a previous generalization do not allow a fully deterministic strong selection limit and propose a new generalization of the Moran process with selection at birth as well as selection at death. 
In \refS{Strong} we perform the strong selection limit analytically for the new process.
In \refS{RPS} we consider the process with selection at birth and death for two player games with three strategies, namely the the Rock--Paper--Scissors game. 
Finally, in \refS{Disc} we conclude and discuss our findings.

\section{Evolutionary Game Dynamics in finite populations}\label{sec:EvDyn}

In this section, we recall some important properties of stochastic evolutionary game 
dynamics in finite populations. For simplicity, we restrict ourselves to birth--death processes in which the number of $A$ players 
can change at most by $\pm 1$ in each update step. 

Let $i$ be the number of $A$  individuals players 
in a population of size $N > 2$. 
The number of $B$ players 
is given by $N - i$. 
The transition probabilities to move from $i$ to $i + 1$ and to $i - 1$ are denoted by $T^+_i(\beta)$ and $T^-_i(\beta)$, respectively. 
The probability to stay in the current state is thus $1 - T^+_i(\beta) - T^-_i(\beta)$. 
These microscopic details do not have to be specified further at this point.
The only requirement is that the expressions for $T^\pm_i(\beta) \neq 0$ are analytic in system size, payoffs and intensity of selection. 
It is also assumed that the population size remains constant. 
Mutations are excluded, such that a strategy 
that is lost will not re-appear in the system. 

In the continuous limit $N\to\infty$, the state of the system $x=i/N$ becomes a continuous variable, strategy $A$ can have any abundance and we recover a deterministic differential equation \cite{helbing:1993aa,helbing:1996aa,schlag:1998aa,traulsen:2005hp,traulsen:2006ab}.
This allows to compute the fixed points of the system. 
There are always fixed points at $x=0$ and $x=1$.
In addition, there can be a third fixed point at $x^\ast=(d-b)/(a-b-c+d)$, which is unstable when $a>c$ and $b<d$ and stable when $a<c$ and $b>d$.

In finite population models, stochasticity does not allow the definition
of fixed points. 
However, the boundaries $i=0$ and $i=N$ are absorbing due to the absence of mutations, 
$T^+_0(\beta) = 0$ and $T^-_N(\beta) = 0$.
For recurrent Markov chains ($T^\pm_i(\beta) > 0$ for $0<i<N$), the system will eventually be absorbed at the boundaries. 
The probability $\phi^A_i(\beta)$ that a given number of $i$ $A$ players will reach the absorbing boundary at $N$ is an important quantity to describe the process. 
In addition to this fixation probability, the unconditional and conditional fixation times, $t_i(\beta)$ and $t^A_i(\beta)$, characterize the stochastic process \cite{antal:2006aa,altrock:2008nj,traulsen:2009bb,cremer:2008aa}.
These two average times are the expectation values of the number of time steps it needs either to reach any homogenous state (all $A$ or all $B$) or 
to reach fixation at all $A$ under the condition that this event occurs. 
In the following we recall recursions for these three quantities which can be solved regardless of the details of the birth--death process. 
Each solution is only based on the microscopic transition probabilities $T^\pm_i(\beta)$.

\subsection{Fixation probability}
 
The probability of fixation $\phi^A_i(\beta)$ describes the probability that a given 
number $i$ of new $A$ mutants in a population of $B$ will reach fixation at all $A$. 
Since the homogenous states are absorbing, we have 
$\phi^{A}_0(\beta) = 0$ as well as $\phi^A_{N}(\beta) = 1$. 
For all the intermediate states we can write a balance equation
for the probability to fixate at all $A$ ($i=N$),
\begin{align}\label{eq:FixProbRec}
\begin{split}	
	0 &= \,T^-_i(\beta)\,\left(\phi_{i-1}^{A}(\beta)-\phi^A_i(\beta)\right)\\
	&+  \, T^+_i(\beta)\,\left(\phi_{i+1}^{A}(\beta)-\phi^A_i(\beta)\right).
\end{split}
\end{align}
With the boundary conditions $\phi_0^{A}(\beta)=0$ and $\phi^A_{N}(\beta)=1$, this can be solved recursively \cite{nowak:2006bo,antal:2006aa}. 
We obtain
\begin{align}\label{eq:FixProbGeneral}
	\phi^A_i(\beta) =\frac
	{1+\sum_{k=1}^{i-1}\prod_{m=1}^{k} \frac{T^-_m(\beta)}{T^+_m(\beta)}}
	{1+\sum_{k=1}^{N-1}\prod_{m=1}^{k}\frac{T^-_m(\beta)}{T^+_m(\beta)}}.
\end{align}
The probability to fixate at the pure state all $B$ starting from $i$ $A$ individuals is given by $\phi^B_i(\beta)=1-\phi^A_i(\beta)$.

\subsection{Unconditional average fixation time}

The average time (measured in elementary time steps) it needs to reach fixation at one of the homogenous states ($i = 0$ or $i = N$) starting with $i$  players of type $A$ is denoted by $t_i(\beta)$. 
Obviously, $t_0(\beta) = 0$ and $t_N(\beta) = 0$. 
The unconditional fixation times also fulfill a balance equation with the transition probabilities describing the rate of change,
\begin{align}\label{eq:UFixTRec}
	t_i(\beta) =1 &+ T^-_i(\beta)\,t_{i-1}(\beta)\nonumber\\
	&+\left( 1 - T ^+_{i}(\beta) - T^-_i(\beta) \right)\,t_i(\beta)\nonumber\\
	&+T^+_i(\beta)\,t_{i+1}(\beta).
\end{align}
This is a recursion equation for the unconditional mean exit times or average times of fixation. 
Its solution for the unconditional average fixation time reads \cite{traulsen:2009bb,goel:1974aa}
\begin{align}\label{eq:UfixTIME}
	t_i(\beta) =& \,\sum\limits_{k=i}^{N-1}\sum\limits_{l=1}^k\frac{1}{T^+_l(\beta)}\prod\limits_{m=l+1}^{k}\frac{T^-_m(\beta)}{T^+_m(\beta)}\nonumber\\ 
	&- t_1(\beta)\sum\limits_{k=i}^{N-1}\prod\limits_{m=1}^{k} \frac{T^-_m(\beta)}{T^+_m(\beta)},\\
	t_1(\beta)=&\phi^A_1(\beta)\sum\limits_{k=1}^{N-1}\sum\limits_{l=1}^k\frac{1}{T^+_l(\beta)}\prod\limits_{m=l+1}^{k}\frac{T^-_m(\beta)}{T^+_m(\beta)}.
\end{align}
Next, we address the time it takes to reach a particular absorbing state.

\subsection{Conditional average fixation time}

Under the condition that the process reaches the absorbing state at all $A$, 
$i=N$, the average time of fixation starting from $i$, is $t^A_i(\beta)$. 
Following \citep{antal:2006aa}, we start from the recursion
\begin{align}\label{eq:CFixTRec}
	\phi^A_i(\beta)\,t^A_i(\beta) =\,\,& T^-_i(\beta)\,\phi^A_{i-1}(\beta)(t^A_{i-1}(\beta)+1)\nonumber\\
	&+\left( 1 - T ^+_{i}(\beta) - T^-_i(\beta) \right)\,\phi^A_i(\beta)(t^A_i(\beta)+1)\nonumber\\
	&+T^+_i(\beta)\,\phi^A_{i+1}(\beta)(t^A_{i+1}(\beta)+1).
\end{align}
The conditional average fixation time is conditioned upon fixation at all $A$, which occurs with probability $\phi^A_i(\beta)$.
Thus, the product of probability and conditional time of fixation appears in the recursion.
There can be a finite expectation value even for vanishing fixation probability, because $\phi^A_i(\beta)t^A_i(\beta)\to0$ does not imply $t^A_i(\beta)\to0$.
For the average fixation time under the condition of absorption at all $A$ we find \citep{antal:2006aa,kampen:1997xg,Fisher1988aa,traulsen:2009bb},
\begin{align}\label{eq:CfixTIME}
	t^A_i(\beta) =& \frac{1}{\phi^A_i(\beta)}\sum\limits_{k=i}^{N-1}\sum\limits_{l=1}^k\frac{\phi^A_l(\beta)}{T^+_l(\beta)}\prod\limits_{m=l+1}^{k}\frac{T^-_m(\beta)}{T^+_m(\beta)}\nonumber\\
			&- t^A_1(\beta)\frac{\phi^A_1(\beta)}{\phi^A_i(\beta)}\sum\limits_{k=i}^{N-1}\prod\limits_{m=1}^{k}\frac{T^-_m(\beta)}{T^+_m(\beta)}, \\
	t^A_1(\beta) =&  \sum\limits_{k=1}^{N-1}\sum\limits_{l=1}^k\frac{\phi^A_l(\beta)}{T^+_l(\beta)}\prod\limits_{m=l+1}^{k}\frac{T^-_m(\beta)}{T^+_m(\beta)}.
\end{align}
There is an analogous expression for the fixation time under the condition that strategy $B$ gets fixed in the population, $t^B_i(\beta)$, \citep{traulsen:2009bb}. 

As the expressions \eq{FixProbGeneral}, \eq{UfixTIME} and \eq{CfixTIME} are functions of the $T^{\pm}_{i}(\beta)$, the study of a strong selection limit has to be performed in the transition probabilities. 
In the next section, we introduce a process with the analytical strong selection limit  $T^+_i(\beta \to \infty)\to1$ and $T^-_i(\beta \to \infty )\to0$ (or the other way around), 
and we show that the resulting dynamics is fully deterministic in this limit.

\section{The Moran process}
\label{sec:Moran}
\subsection{Selection at birth and random death}

A standard model for evolutionary dynamics in finite populations is the frequency 
dependent Moran process \citep{nowak:2004pw}. This process incorporates the following 
steps: 
An individual is selected at random, but proportional to its fitness. 
This individual produces identical offspring. 
The offspring replaces an individual randomly selected for death. 
Fitness $f$ is a convex combination of the average payoffs from the game, $\pi_i^A$ and $\pi_i^B$, and a background fitness, which is usually set to one. 
Thus, we have $f_i^A(\beta) = 1 - \beta + \beta \,\pi_i^A$ and $f^{B}_{i}(\beta) = 1 - \beta + \beta\,\pi_i^B$. 
The quantity $0\leq\beta\leq\beta_{\text{max}}$ determines the intensity of selection. 
The transition probabilities of the Moran process are thus given by 
\begin{subequations}
\begin{align}
T^+_i(\beta)
&=
\underbrace{
\frac{i \, f_i^A(\beta) }{i f_i^A(\beta) +(N-i) f_i^B(\beta)}
}_{\hbox{\footnotesize Selection at birth}}
\times
\underbrace{
\frac{N-i }{N},
}_{\hbox{\footnotesize Random death}}\label{eq:TMoranPlus}
\\
T^-_i(\beta)
&=
\underbrace{
\frac{(N-i) f_i^B(\beta) }{i f_i^A(\beta) +(N-i) f_i^B(\beta)}
}_{\hbox{\footnotesize Selection at birth}}
\times
\underbrace{
\frac{i }{N}.
}_{\hbox{\footnotesize Random death}} \label{eq:TMoranMinus}
\end{align}
\end{subequations}
Selecting proportional to fitness implies that fitness is positive. 
Thus, for payoff matrices with negative entries, the intensity of selection $\beta$ cannot exceed a threshold $\beta_{\text{max}}$. 
This process does not have a generic deterministic limit with arbitrarily strong selection intensity and remains stochastic with random death.

A possibility to extend the Moran process to higher intensities of selection is to 
choose fitness as an exponential function of the payoff, i.e.\ $f_i^A(\beta) = \exp(+\beta\,\pi_i^A)$ 
and $f_i^B(\beta) =\exp(+\beta\,\pi_i^B)$ \citep{prugel:1994hb,traulsen:2008aa}. 
Now, the intensity of selection $\beta$ can be any positive number. 
For $\beta\to\infty$, the fitter individual is always selected for reproduction, compare equations \eq{TMoranPlus} and \eq{TMoranMinus}. 
The direction of the process becomes deterministic. 
But due to random death, the system can remain longer or shorter in a particular state. 
Thus, the process remains stochastic in what concerns the times to fixation.
With two strategies, we have a semi deterministic process with deterministic direction and stochastic speed \cite{traulsen:2007cc}.
If there are more than two strategies, random death can also change the composition of the less fit types in the population. 
This can affect the direction of selection, as the fittest type can change due to frequency dependent selection.

\subsection{Selection at birth and death}\label{sec:BDprocess}

Here, we introduce a birth--death process that recovers the usual results for weak selection, 
but also leads to fully deterministic asymptotic behavior for strong selection.
The process has deterministic microscopic dynamics if the $T_i^\pm(\beta)$ are zero or one.
As in the standard Moran process, we assume that selection at birth is proportional to fitness. 
In addition to producing less offspring, individuals with a lower fitness now have a higher probability to die. 
A simple way to incorporate this is to select at death proportional to inverse fitness. 
To ensure that fitness is a positive number, we follow the approach discussed above and 
define fitness as an exponential function of the payoff. 
This leads to the transition probabilities 
\begin{subequations}
\begin{align}
T^+_i(\beta)
=\, &
\underbrace{
\frac{i\e^{+\beta_b\,\pi_i^A}}{i\e^{+\beta_b \pi_i^A}+(N-i)\e^{+\beta_b\, \pi_i^B}}
}_{\hbox{\footnotesize Selection at birth}}\nonumber\\
&\times
\underbrace{
\frac{(N-i)\e^{-\beta_d \,\pi_i^B}}{i\e^{-\beta_d\, \pi_i^A}+(N-i)\e^{-\beta_d\, \pi_i^B}}
}_{\hbox{\footnotesize Selection at death}}\label{eq:TBDPlus},
\\
T^-_i(\beta)
= \, &
\underbrace{
\frac{(N-i)\e^{+\beta_b\,\pi_i^B}}{i\e^{+\beta_b \,\pi_i^A}+(N-i)\e^{+\beta_b \,\pi_i^B}}
}_{\hbox{\footnotesize Selection at birth}}\nonumber\\
&\times
\underbrace{
\frac{ie^{-\beta_d \,\pi_i^A}}{i\e^{-\beta_d \,\pi_i^A}+(N-i)\e^{-\beta_d \,\pi_i^B}}
}_{\hbox{\footnotesize Selection at death}}\label{eq:TBDMinus}.
\end{align}
\end{subequations}
$\beta_b$ is the intensity of selection at birth and $\beta_d$ is the intensity of selection at death.
For $\beta_d=0$, we recover the process discussed in \refS{Moran}. 
It is known that under weak selection many birth--death processes have the same general properties \citep{fudenberg:2006fu,lessard:2007aa, taylor:2006jt}. 
Especially, for $\beta_{b,d}\ll1$ the behavior of the Moran process is recovered \citep{traulsen:2007cc, altrock:2008nj}. 

For simplicity, we assume $\beta=\beta_{b}=\beta_{d}$ in the following. 
The transition probabilities can be written as
\begin{subequations}
\begin{align}
T^+_i(\beta)
&=
\frac{i}{i+(N-i)\e^{-\beta\, \Delta\pi_i}}
\frac{N-i}{i\,\e^{-\beta\, \Delta\pi_i}+N-i}\label{eq:Tstrong01a},
\\
T^-_i(\beta)
&=
\frac{N-i}{i\,\e^{+\beta\, \Delta\pi_i}+N-i}
\frac{i}{i+(N-i)\e^{+\beta\, \Delta\pi_i}}\label{eq:Tstrong01b}.
\end{align}
\end{subequations}
Thus, as far as the payoffs are concerned, the transition probabilities only depend on the difference between the frequency dependent average payoffs, $\Delta\pi_i=\pi_i^A-\pi_i^B$. 

The case of $\Delta\pi_i=0$ is a form of neutral selection.
In this case, the transition probabilities are $T_i^\pm(\beta)\equiv(N-i)/N^2$, for arbitrary $\beta$. 
Note that for neutral selection, moving into one direction is equally probable as moving into the other $T^+_i(0)=T^-_i(0)$.
But the probability to leave a given interior state changes with $i$,
$T_i^\pm(0)\neq T_j^\pm(0)$ for $ i\neq j $. 

For arbitrary $\beta$, the ratio of the transition probabilities reduces to an exponential function
of the payoff difference,
\begin{align}
\frac{T^-_i(\beta)}{T^+_i(\beta)} = e^{- 2 \beta \Delta\pi_i}. 
\end{align}
Hence, the fixation probabilities of the process can be approximated with the closed
expressions derived in \cite{traulsen:2006bb}, after rescaling the intensity of selection
by a factor of 2. From this, it is clear that the usual weak selection behavior is recovered.

\section{Strong selection}
\label{sec:Strong}

For strong selection, $\beta \to \infty$, the asymptotic behavior of the transition probabilities depends only on the sign of the payoff difference.
We focus on the generic cases $\Delta\pi_i\lessgtr0$ to discuss this limit.
The limiting cases can be obtained from equations \eq{Tstrong01a} and \eq{Tstrong01b} and yield
\begin{align}\label{eq:Tstrong02a}
	\lim_{\beta\to\infty}\,T^+_i(\beta) =
	\begin{cases}
			\,0 \hspace{0.2cm}\,&\text{for}\,\,\,\Delta\pi_i<0\vspace{0.25cm}\\
			\,1\hspace{0.2cm}\,&\text{for}\,\,\,\Delta\pi_i>0
	\end{cases},
\end{align}
as well as
\begin{align}\label{eq:Tstrong02b}
	\lim_{\beta\to\infty}\,T^-_i(\beta) =
	\begin{cases}
			\,1 \hspace{0.2cm}\,&\text{for}\,\,\,\Delta\pi_i<0\vspace{0.25cm}\\
			\,0\hspace{0.2cm}\,&\text{for}\,\,\,\Delta\pi_i>0
	\end{cases}.
\end{align}
Since $\lim_{\beta\to\infty}(T_i^+(\beta)+T_i^-(\beta))=1$, the probability to stay in the state $i$ ($0<i<N$) vanishes for $\beta\to\infty$ and nontrivial payoff  difference, $\Delta\pi_i\neq0$.
Thus, we have a fully deterministic process for arbitrary population size.
With this, we consider the fixation probability $\phi^A_i(\beta)$ and average fixation times $t_i(\beta)$ and $t^A_i(\beta)$ in the limiting case of strong selection.
In the following let $\phi_i^A(\infty)$ as well as $t_i(\infty)$ and $t_i^A(\infty)$ denote the finite asymptotic ($\beta\to\infty$) values of the fixation probability and times. 
We identify them in terms of the initial frequency $i$, depending on the average payoff difference for the process introduced above under strong selection.

\subsection{Fixation Probability}\label{ssec:Strong01}

Starting with equation \eq{FixProbRec} and inserting the limiting cases of $T_i^\pm(\beta)$ leads to
\begin{align}\label{eq:FixProbL01}
	\phi_i^A(\infty)=&\,\left(\lim_{\beta\to\infty}T^-_i(\beta)\right) \phi^A_{i-1}(\infty)\nonumber\\
	&+\left(\lim_{\beta\to\infty}T^+_i(\beta)\right) \phi^A_{i+1}(\infty).
\end{align}
That is, in the strong selection limit we have a very simple recursion for the asymptotic value of the fixation probability, 
depending on the sign of the payoff difference $\Delta\pi_i$.
When strategy $A$ dominates strategy $B$ ($a>c$ and $b>d$), we have $\Delta\pi_i>0$. 
For  $0 < i < N$, this yields $\lim_{\beta\to\infty}T^-_i(\beta) = 0$, and
$\lim_{\beta\to\infty}T^+_i(\beta) = 1$,
which results in $\phi_i^A(\infty) = 1-\delta_{0,i}$. 
In other words, the probability to reach the state with all $B$ individuals is zero, except if there are no $A$ individuals initially. 
Equivalently, for dominance of strategy $B$ we obtain  $\phi_i^A(\infty) = \delta_{N,i}$.
More interesting cases are found when $\Delta\pi_i$ changes its sign, which occurs at $i^\ast=(N(d-b)+a-d)/(a-b-c+d)$ when $a>c$ and $d>b$ or when $a<c$ and $d<b$.
These are two important classes of games: coordination games and coexistence games. 

\begin{figure}[h]
	\includegraphics[width=0.475\textwidth,angle=0]{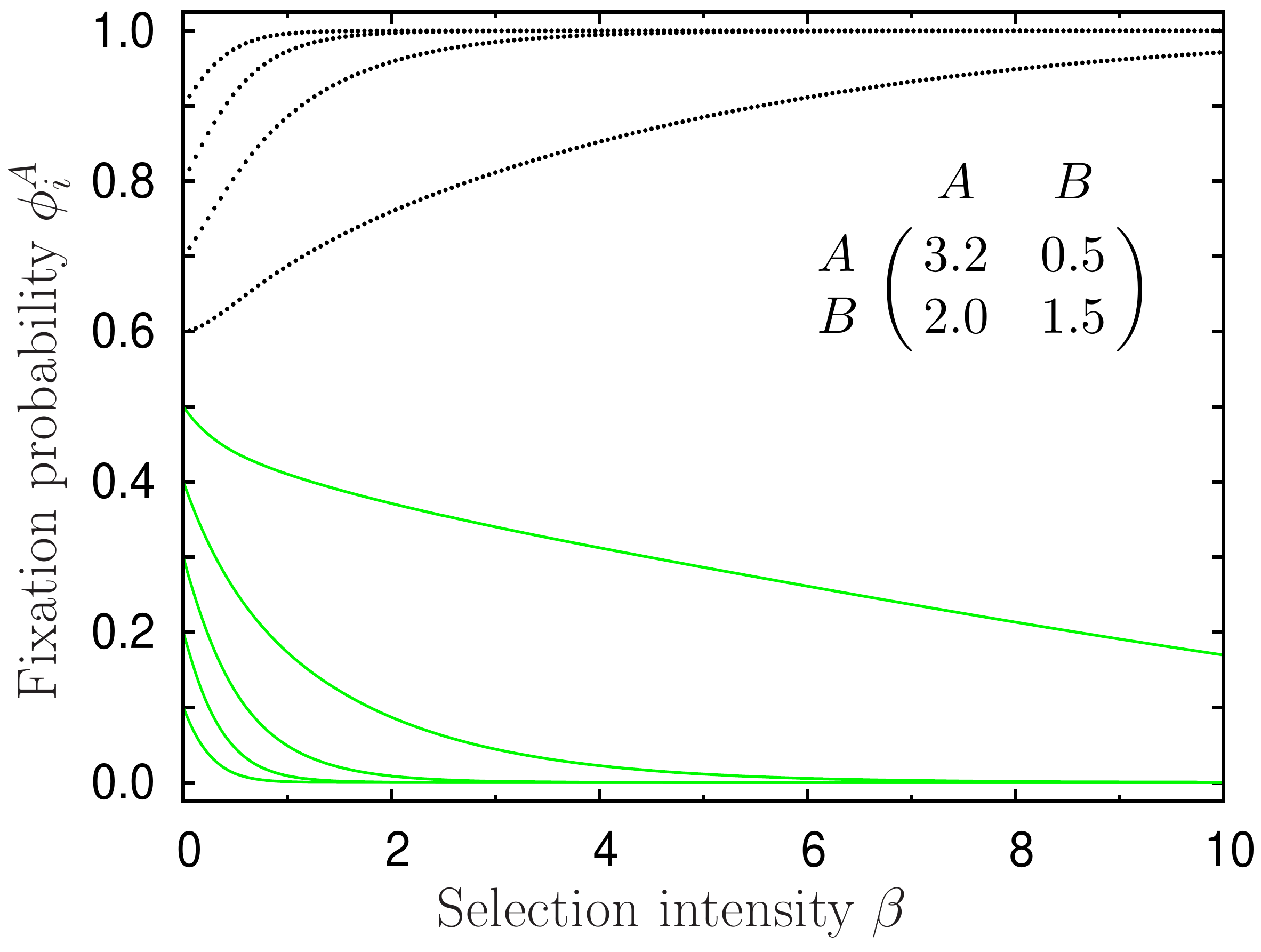}
\caption{
The fixation probability as a function of selection intensity for a coordination game ($a>c$, $b<d$), given by Eq.~\eq{FixProbGeneral}, with selection at birth and death, Eqs.~\eq{Tstrong01a} and \eq{Tstrong01b}. 
From the payoff matrix given in the figure, we obtain $i^\ast = \frac{17}{22} + \frac{5}{11}N$, which gives $i^\ast \approx5.32$ for our numerical example with $N=10$. 
With increasing selection intensity $\beta$, for any $l<i^\ast$ we have $\phi_l^A(\beta)\to0$ (straight lines), 
whereas for any $m>i^\ast$ we have $\phi_m^A(\beta)\to1$ (dotted lines). 
The $\phi^A_i(\beta)$ for each $i$ can be identified via its neutral value $\phi^{A}_i(0)=\frac{i}{N}$. 
}
\label{fig:FIG1}
\end{figure}

Let us first focus on coordination games ($a>c$ and $d>b$). 
In these games, the threshold value $i^\ast$ cannot be crossed for infinitely large systems with deterministic dynamics \cite{skyrms:2003aa}.  
For finite systems under strong selection, we observe something similar.
The payoff difference $\Delta\pi_i$ changes sign from negative to positive and selection points always away from $i^\ast$ toward the boundaries. 
Depending on the initial condition $i\gtrless i^\ast$, the fixation probability $\phi^A_i(\infty)$ is one or zero.
\begin{itemize}
\item[(i)] If $i<i^\ast$, $\Delta\pi_i$ is negative and with Eqs.\ \eq{Tstrong02a} and \eq{Tstrong02b} we have $\phi_i^A(\infty)= \phi^A_{i-1}(\infty)$. 
We start the recursion for the fixation probabilities with $\phi^A_1(\infty)=\phi^A_0(\infty)=0$.
This yields $\phi^A_{i<i^\ast}(\infty)=0$.
\item[(ii)] For $i>i^\ast$, $\Delta\pi_i$ is positive and Eqs.\ \eq{Tstrong02a} and \eq{Tstrong02b} yield the recursion $\phi_i^A(\infty)=\phi^A_{i+1}(\infty)$. 
Starting with the maximal $i$ we obtain $\phi^A_{N-1}(\infty)=\phi^A_N(\infty)=1$
and thus $\phi^A_{i>i^\ast}(\infty)=1$. 
\item[(iii)] If $i^\ast$ happens to be an integer value and the system starts there, the first step has equal probabilities, $T^+_{i^\ast}(\beta)=T^-_{i^\ast}(\beta)=\frac{1}{2}$.
This leads to $\phi^A_{i^\ast}(\infty)=\frac{1}{2}$.
\end{itemize}
In summary, for the fixation probability we find
\begin{align}\label{eq:FixProbL02}
	\,\phi_i^A(\infty) =
	\begin{cases}
		\,0\hspace{0.2cm}\,&\text{for}\,\,\,i<i^\ast \vspace{0.25cm}\\
		\,\frac{1}{2}\hspace{0.2cm}\,&\text{for}\,\,\,i=i^\ast \vspace{0.25cm}\\
		\,1\hspace{0.2cm}\,&\text{for}\,\,\,i>i^\ast
	\end{cases}.
\end{align} 
This is clearly what is to be expected because $A$ is selected for $i> i^{\ast}$ and $B$ is selected for $i< i^{\ast}$, \fig{FIG1}. 
Dominance of strategy $A$ can be seen as a special case of coexistence with $i^{\ast}<0$. 

Next, we consider coexistence games with $a<c$ and $d<b$. 
In this case, $\Delta\pi_i$ changes sign from positive to negative and selection points always away from the boundaries toward $i^\ast$. For strong selection, the system gets trapped and fixation never occurs. 
\begin{itemize}
\item[(i)] If $i^\ast$ is an integer, the system switches from $i^\ast$ to $i^\ast \pm 1$ with equal probability. 
From $i^\ast \pm 1$, it always returns to $i^{\ast}$. 
\item[(ii)] If $i^\ast$ is not an integer, we observe deterministic flipping between the two neighboring states states $i_1<i^\ast$, $i_2>i^\ast$. 
\end{itemize}
Since fixation never occurs in coexistence games, it does not make sense to compute the asymptotic value of the fixation probability.
Formally, the probability to get absorbed in all $A$ converges to $1$ if $i^{\ast}>N/2$ and to $0$ otherwise. 
However, it turns out that the fixation times diverge. 

\subsection{Unconditional average fixation time}\label{ssec:Strong02}
 
The average time it takes for $i$ $A$ players to either become extinct or take over the population, $t_i(\beta)$, can be found by solving equation \eq{UFixTRec} recursively. 
To examine the limit of strong selection, we perform the limit on both sides of the balance equation, assuming that there exists an asymptotic value $t_i(\infty)$ of the unconditional fixation time.
With the previous analysis of the transition probabilities, this leads to
\begin{align}\label{eq:UFixTL01}
	t_i(\infty) = 1&+\left( \lim_{\beta\to\infty}T^-_i(\beta)  \right)\,t_{i-1}(\infty)\nonumber\\
	&+\left( \lim_{\beta\to\infty}T^+_i(\beta)  \right)\,t_{i+1}(\infty).
\end{align}
This strong selection recursion has to be analyzed for the two different cases of behavior at the threshold $i^\ast$, 
coordination and coexistence.
Again, we first examine the coordination game ($a>c$ and $d>b$). 
\begin{itemize}
\item[(i)] For $i<i^\ast$, the payoff difference is negative, $\Delta\pi_i<0$. 
The recursion amounts to $t_i(\infty)=1+t_{i-1}(\infty)$, because $T^+(i)\to0$ whereas $T_i^-(\beta)\to1$.
With $t_0(\infty)=0$ at the boundary we have $t_1(\infty)=1$, $t_2(\infty)=1+t_1(\infty)=2$, and eventually $t_i(\infty)=i$.
\item[(ii)] For $i>i^\ast$ the payoff difference is $\Delta\pi_i>0$. The recursion from \eq{UFixTL01} is $t_i(\infty)=1+t_{i+1}(\infty)$. 
The transition probabilities $T^{\pm}(i)$ behave exactly in the opposite way as before.
The time starting from next to the absorbing boundary is $t_{N-1}(\infty)=1+t_N(\infty)=1$. 
Hence, we have $t_{N-k}(\infty)=k$, or with $k=N-i$, the uncondtional averag time is $t_i(\infty)=N-i$.
\item[(iii)] If the threshold is an integer and the system is initiated there the number of steps needed to fixate is $i^\ast$ or $N-i^\ast$ with equal probability.
Thus, we can compute the average time with the previous findings, $t_{i^\ast}(\infty)=1/2(i^\ast+N-i^\ast)=N/2$.
\end{itemize}
In summary, depending on the the starting point $i^{\ast}$, the asymptotic value for the unconditional average fixation time in a coordination game is
\begin{align}\label{eq:UFixTL02}
	t_i(\infty) =
	\begin{cases}
		\,i\hspace{0.2cm}\,&\text{for}\,\,\,i<i^\ast,\vspace{0.25cm}\\
		\,\frac{N}{2}\hspace{0.2cm}\,&\text{for}\,\,\,i=i^\ast,\vspace{0.25cm}\\
		\,N-i\hspace{0.2cm}\,&\text{for}\,\,\,i>i^\ast.
	\end{cases}
\end{align} 
In the strong selection limit $t_i(\infty)$ converges to the distance between initial state and the final state, as expected from deterministic motion. 

\begin{figure}[h,t]
	\includegraphics[width=0.475\textwidth,angle=0]{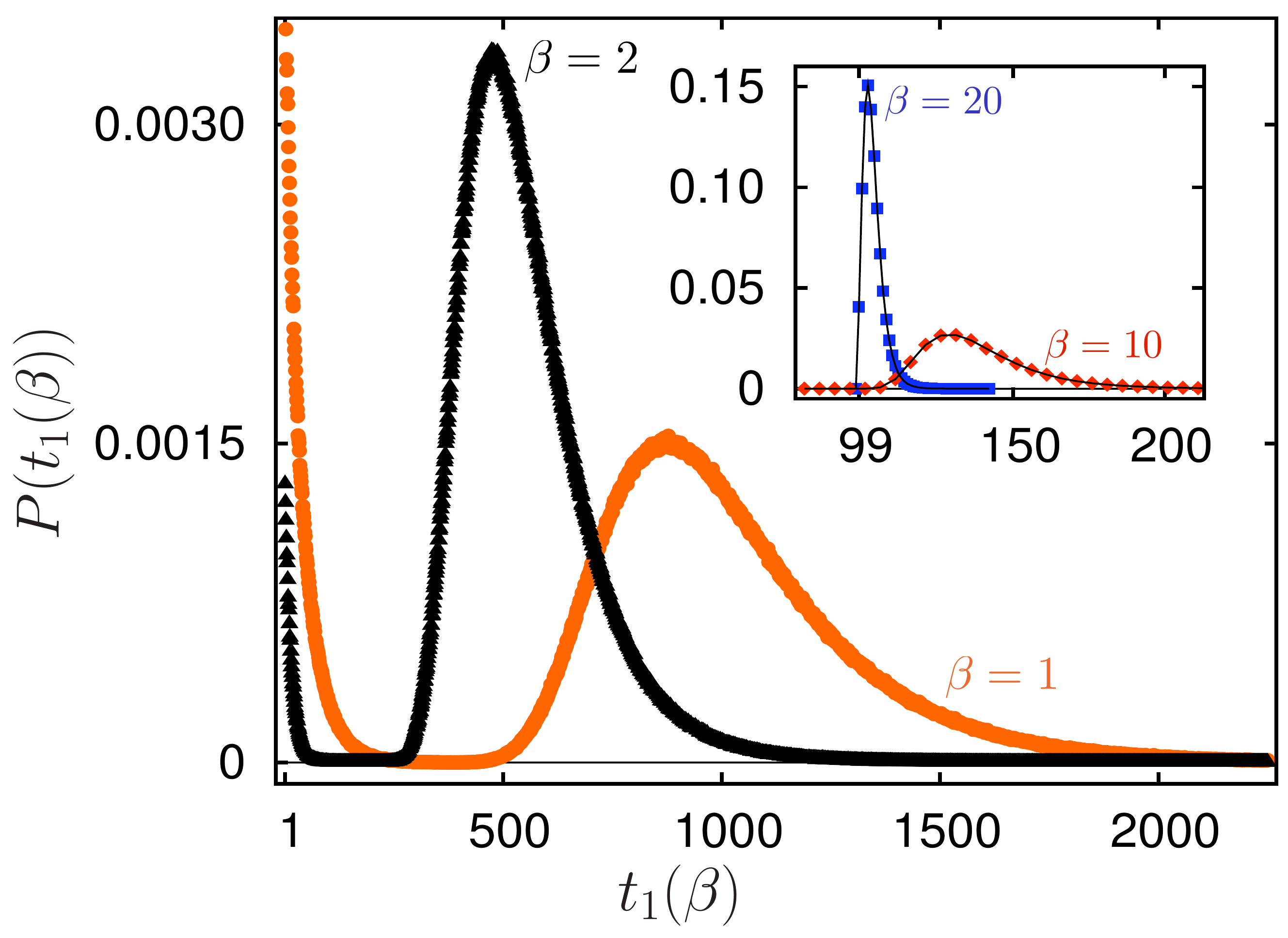}
\caption{ 
Probability distribution of the unconditional fixation time  (measured in elementary time steps) of a single $A$ player in a population of $N-1$ dominated $B$ players.
For $\beta=1$ and $\beta=2$, the distribution has two peaks corresponding to the two absorbing boundaries. 
For stronger selection (inset), the probability that the advantageous $A$ individual goes extinct becomes small and 
fixation takes at least $N-1$ time steps. In this case, the distribution becomes single peaked. 
For $\beta \to \infty$, the distribution converges to a delta peak at $t_1(\infty)=N-1$ (payoff matrix $a=2.2$, $b=1.5$, $c=2$, $d=0.5$, population size $N=100$,
histograms obtained from $10^7$ realizations. 
Lines are guides to the eye). 
}
\label{fig:FIG2}
\end{figure}

We can also infer the dynamics for games in which one strategy dominates.
When strategy $A$ dominates, we can formally set $i^{\ast} <0$ and obtain $t_i(\infty)=N-i$,
cf.~\fig{FIG2}. 
When strategy $B$ dominates, the equivalent procedure yields $t_i(\infty)=i$.

In case of a coexistence game, the system gets trapped around $i^\ast$ and cannot reach the absorbing boundaries. As expected, the recursions lead to $t_i(\infty)\to\infty$.

\subsection{Conditional average fixation time}\label{ssec:Strong03}
 
If $i$ $A$ players take over the population, the asymptotic fixation time under this condition, $t_i^A(\infty)$, can be obtained by solving the balance equation \eq{CFixTRec} recursively. 
However, this situation is more complex as we have to consider the 
fixation probability and the conditional fixation time in a combined way.
Introducing the asymptotic value $\theta_i^A(\infty)=\phi_i^A(\infty)\,t_i^A(\infty)$, the recursion \eq{CFixTRec} yields
\begin{align}\label{eq:CFixTL01}
	\theta_i^A(\infty) =\,& \left( \lim_{\beta\to\infty}T^-_i(\beta) \right)(\theta_{i-1}^A(\infty)+\phi^A_{i-1}(\infty))\nonumber\\
	&+\left( \lim_{\beta\to\infty}T^+_i(\beta) \right)(\theta_{i+1}^A(\infty)
	+\phi^A_{i+1}(\infty)).
\end{align}
The formulation of a similar equation for $\theta_i^B(\infty)=\left(1-\phi_i^A(\infty)\right)\,t^{B}_i(\infty)$ is straightforward. 
Both are analyzed regarding the different behavior at either side of the threshold $i^\ast$.

\begin{figure}[h]
	\includegraphics[width=0.475\textwidth,angle=0]{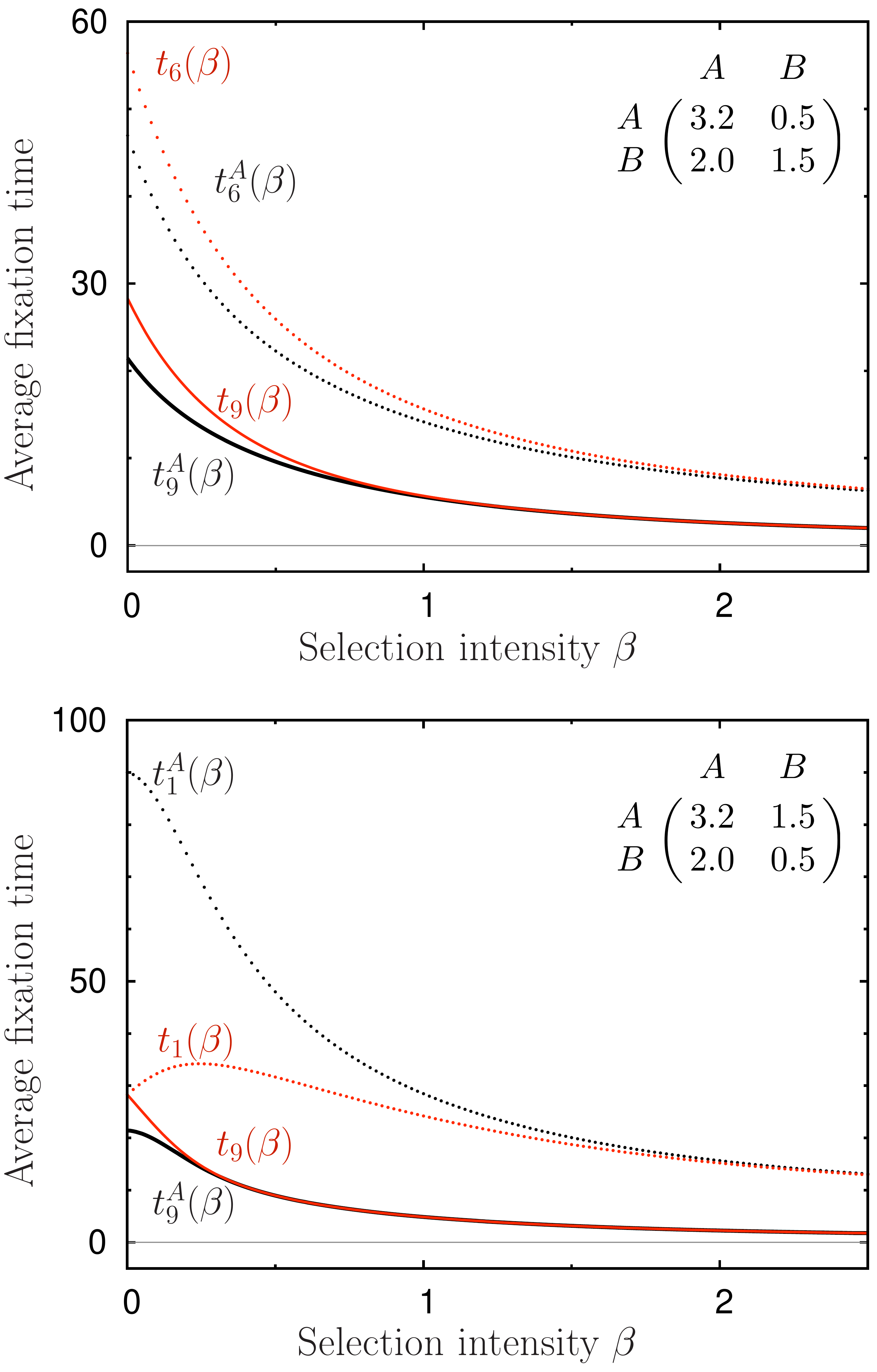}
\caption{
Unconditional (red) and conditional (black) average fixation times  (measured in elementary time steps) as a function of the intensity of selection for a coordination game (top) and a game in which $A$ dominates (bottom).  
The payoff matrices of the games are given in the figures, the population size is $N=10$. 
Top: In a coordination game ($a>c$, $b<d$), the conditional and unconditional fixation times, $t_i(\beta)$ and $t_i^A(\beta)$ converge to $N-i$ for $\beta \to \infty$ if initially  more than $i^{\ast} =117/22\approx5.32$ individuals play $A$, compare Eq.~\eq{UFixTL02}. 
The lines show the initials states $i=6$ (dotted lines) and $i=9$ (full lines). 
Bottom: When $A$ dominates $B$ ($a>c$, $b>d$), the unconditional fixation time $t_1(\beta)$ first increases with $\beta$. 
For any initial condition $i$, $t_i(\beta)$ and $t_i^A(\beta)$ converge to $N-i$ in the limit of strong selection.  
The initials states are $i=1$ (dotted lines) and $i=9$ (straight lines). 
}\label{fig:FIG3}
\end{figure}

For the coordination game the system reaches the absorbing boundaries after a finite time.
\begin{itemize}
\item[(i)] If $i>i^\ast$,  the system fixates at $i=N$ with probability $\phi^A_i(\infty)=1$.
Thus, $\theta_i^A(\infty)=t_i^A(\infty)$ and we recover the same recursion as for the unconditional
fixation time. 
This yields $ t_i^A(\infty) = t_i(\infty) = N-i$, see \fig{FIG3}. 
\item[(ii)] If $i<i^\ast$, the system fixates at $i=0$, $\phi^A_i(\infty)=0$.
Thus, we cannot formulate a meaningful recursion for $t_i^A(\infty)$.
In this case we observe $T^-_i(\beta)\to1$, $T^+_i(\beta)\to0$ and we can only make a statement for $t_i^B(\infty)$,
which results in $t_i^{B}(\beta) = i$.
\item[(iii)] If $i=i^\ast$ is an integer, the system is not fully deterministic as fixation of $A$ and fixation of $B$ are observed with equal probability $\frac{1}{2}$.
In this case, we obtain 
$t_{i^\ast}^A(\infty) = N-i^{\ast}$  
and
$t_{i^\ast}^B(\infty) = i^{\ast}$. 
\end{itemize}
In a regime where $A$ always performs better than $B$ the unconditional fixation time $t_i^A(\infty)$ is equal to the conditional fixation time $t_i(\infty)$, see \fig{FIG3}. 
Equivalently, when $B$ always performs better, we have $t^B_i(\infty)=t_i(\infty)$. 

For a coexistence game, the system does not reach any of the boundaries but is always dragged toward $i^\ast$, as discussed before.
The recursion \eq{CFixTL01} for $t_i^A(\infty)$ or its equivalent for $t_i^{B}(\infty)$
are not meaningful here because they contain the fixation probabilities.
However, in this case all fixation times diverge with $\beta$.

In this section, we have derived asymptotic values for the birth--death process with selection at birth and selection at death. 
We have identified the underlying games that lead to a fully deterministic process in the limit of strong selection. 
As we have seen, the difference of the average payoffs $\Delta\pi_i$ as a function of the the relative abundance of type $A$ plays an important role.

\section{Games with three strategies}\label{sec:RPS}

Here, we demonstrate that the process we have introduced above
 leads to rather simple and often deterministic dynamics. 
We focus on games with two players and three strategies with cyclic dominance \cite{sinervo:1996le,kerr:2002xg,czaran:2002ya,szabo:2004ww,szolnoki:2005wq,reichenbach:2006aa,perc:2007bb,sandholm:2007bo,claussen:2008aa,peltomaki:2008ls}. 
Cyclic dominance among three strategies corresponds to Rock--Paper--Scissors games, where each strategy can be beaten by another one: 
Rock crushes Scissors, Scissors cut Paper and Paper wraps rock.
In general, the payoff for winning does not have to be equal to the payoff for losing,
which leads to non--zero--sum games.
For simplicity, we set the payoff for a tie to zero. 
Setting the winners payoff to one and the losers payoff to $-s \leq 0$, the $3\times3$ payoff matrix reads
\begin{align}\label{eq:RPSmatrix}
\bordermatrix{
  & R & P & S \cr
R & 0 & -s & 1\cr
P & 1 & 0 & -s \cr
S & -s & 1 & 0 \cr}.
\end{align}
For infinite populations, the state of the system is defined by the frequencies of the three strategies, $x_R$, $x_P$, and $x_S$. 
Thus, the state space is the simplex  $S_3\subset\mathbb R^2$, an equilateral triangle between the three states all $R$, all $P$, and all $S$.
Apart from the three trivial equilibria, the replicator dynamics has an
interior equilibrium at $ (x_R, x_P, x_S)^{\ast}=(\frac{1}{3}, \frac{1}{3},\frac{1}{3})$, which follows from the symmetry of the system. 
The quantity $s$ determines wether the interior equilibrium is asymptotically stable (the system spirals inwards towards the interior fixed point for $s<1$) or unstable (the system spirals out toward a heteroclinic cycle along the boundaries for $s>1$). 
In the zero--sum game with $s=1$, the system oscillates around the interior equilibrium with the Hamiltonian $-x_Rx_Px_S$ being a constant of motion \citep{hofbauer:1998mm, nowak:2006bo}.
\begin{figure}[h]
\begin{center}
	\includegraphics[width=0.45\textwidth,angle=0]{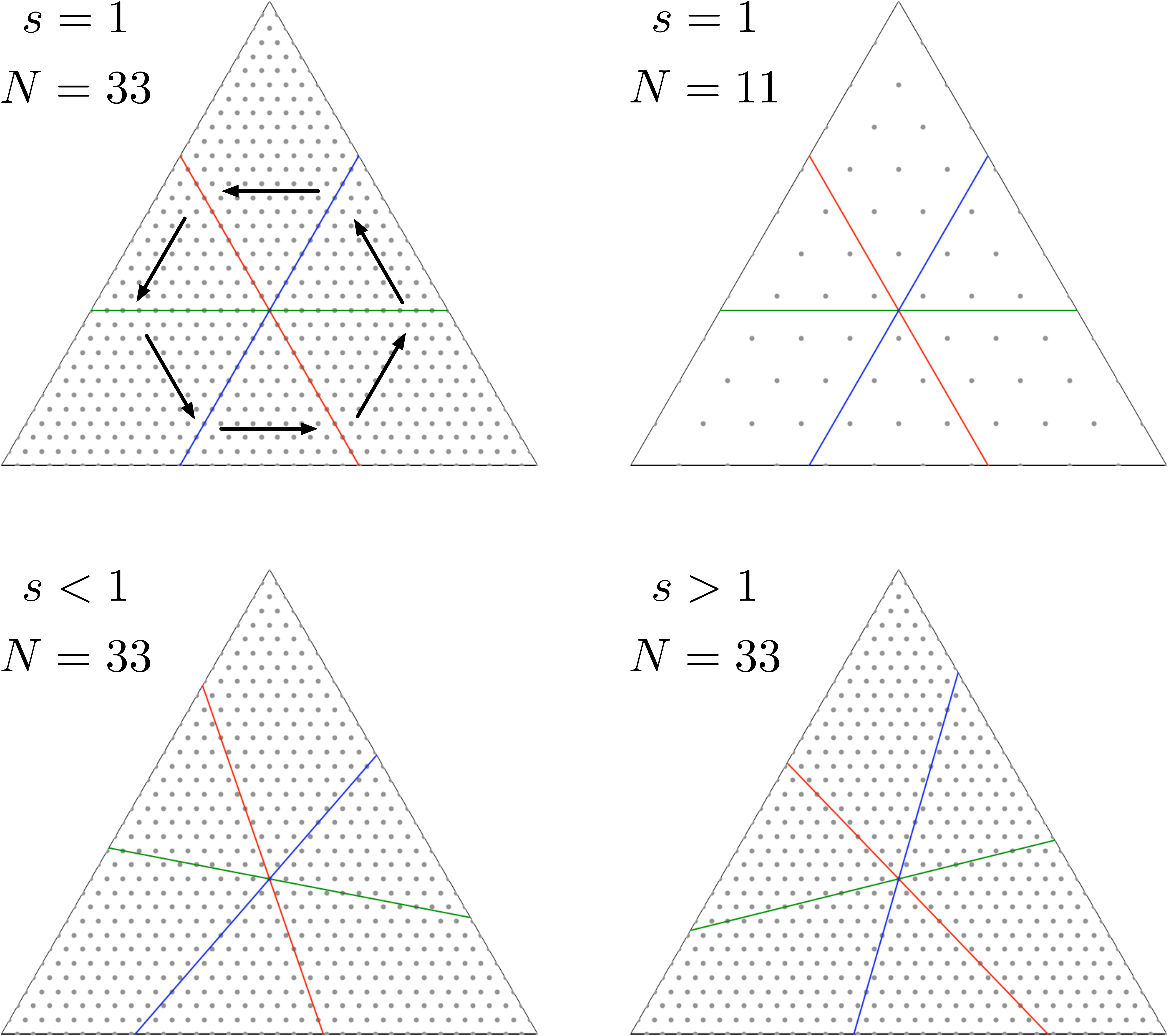}
	\end{center}
\caption{ 
The discrete simplex of the Rock--Paper--Scissors game for the different ranges of $s$. 
The three strategies are arranged in such a way that cyclic dominance is counter clockwise. 
For $N=33$ and $s=1$ (top left), the three lines of equal payoffs defined by Eqs. \eq{RPSPayEq}
are parallel to the boundaries of the simplex. Changing the population size to $N=11$ (top right) does not
change these lines, but only the state space of the system indicated by dots. For example, the system can no longer access the center of the simplex. 
With decreasing $s$, the three lines of equal payoffs are rotated clockwise (bottom left, $s=0.5$). 
With increasing $s$, these lines are rotated counterclockwise (bottom right, $s=2.5$).
Only in special cases, the three lines of equal payoffs defined by Eqs. \eq{RPSPayEq}
intersect with the possible states of the system. 
The arrows in the upper left figure indicate the direction of selection as it is induced by the cyclic dominance of the three strategies.
}\label{fig:FIG4}
\end{figure}
In finite, well--mixed populations, the state space is only a subset of $(N+1)(N+2)/2$ states within the simplex  $S_3$, cf. \fig{FIG4}.
Moreover, the dynamics is typically stochastic. 
Properties such as the average drift for a Moran process or the average time to reach the absorbing the boundaries are attainable for weak selection \citep{reichenbach:2006aa,claussen:2008aa}.
To analyze the strong selection limit, we adopt the evolutionary process with selection at birth and death discussed above for $2\times2$ games. 
Starting with the payoffs from \eq{RPSmatrix} in a well--mixed population the average payoffs read
\begin{align}
	\pi_R
	&= i_S - i_P\,s\label{eq:RPS01},\\
	\pi_P
	&= i_R -i_S\,s\label{eq:RPS02},\\
	\pi_S
	&= i_P -i_R\,s\label{eq:RPS03},
\end{align}
where $i_R$, $i_P$, and $i_S=N-i_R-i_P$ are the number of individuals playing Rock, Paper, or Scissors in a population of size $N$, respectively. 
Individuals are selected proportional to fitness at birth and proportional to inverse fitness at death, 
both with the intensity of selection $\beta$. 
The fitness of strategy $X$ is given by $f_X=\exp [ + \beta\,\pi_X ]$. 
The dynamics on the discrete finite set of states is governed by six transition probabilities in each state. 
The transition probabilities in each state $(i_R,i_P,i_S)$ to change to one of the six neighboring states are thus given by
\begin{align}\label{eq:RPST01}
T_{Y\to X}(i_R,i_P,i_S) \,=\,
\underbrace{
	\frac{i_X \,f_X}{\sum_{Z} i_{Z}f_{Z}}
}_{\hbox{\footnotesize birth}}
\times
\underbrace{
	\frac{i_Y \,f_Y^{-1}}{\sum_{Z} i_{Z}f_{Z}^{-1}}
\,
}_{\hbox{\footnotesize death}},
\end{align}
where $X$, $Y$, and $Z$ stand for $R$, $P$, or $S$. 
Note that the probability to stay in the given state is given by $T_{R\to R}(i_R,i_P,i_S)+T_{P\to P}(i_R,i_P,i_S)+T_{S\to S}(i_R,i_P,i_S)$. 
For strong selection, $\beta\to\infty$, the system moves from each state into one direction with probability one unless two payoffs are identical. 
In the following we address how this direction depends on the payoffs in the state and on the paprameter $s$. 

Let us first assume that for given state $(i_R,i_P,i_S)$, we have the unique ordering of the average payoffs from \eq{RPS01}--\eq{RPS03}. 
Let $\pi_1$ denote the largest and $\pi_3$ denote the lowest value, \ie $\pi_1>\pi_2>\pi_3$.
The number of individuals playing the according strategies  
can be denoted as $i_1$, $i_2$, and $i_3$.
For $T_{3\to 1}$, we have 
\begin{align}
	T_{3\to 1} =\,& \frac{i_1\e^{\beta\pi_1}}{i_1\e^{\beta\pi_1}+i_2\e^{\beta\pi_2}+i_3\e^{\beta\pi_3}}\nonumber\\
	&\times\frac{i_3\e^{-\beta\pi_3}}{i_1\e^{-\beta\pi_1}+i_2\e^{-\beta\pi_2}+i_3\e^{-\beta\pi_3}}\label{eq:RPSTd}\\
				=\,&\frac{i_1}{i_1+i_2\e^{-\beta(\pi_1-\pi_2)}+i_3\e^{-\beta(\pi_1-\pi_3)}}\nonumber\\
	&\times\frac{i_3}{i_1\e^{-\beta(\pi_1-\pi_3)}+i_2\e^{-\beta(\pi_2-\pi_3)}+i_3}\label{eq:RPSTe},
\end{align}
where $\pi_1-\pi_2 > 0$,
$\pi_1-\pi_3 >0$, as well as 
$\pi_2-\pi_3>0$.
For $\beta\to\infty$, this leads to 
\begin{align}\label{eq:RPSLim01}
	\lim_{\beta\to\infty}T_{3\to 1} = 1.
\end{align}
All the other transition probabilities vanish. 
In each reproductive event, an individual with the largest payoff replaces an individual with the smallest payoff. 
This holds for any unique ordering of the three payoffs.

If the payoffs are not in unique order, that is, if two or more payoffs are equal, at least two probabilities become non--trivial. 
This yields the following three scenarios:
\begin{itemize}
\item[(i)] For $\pi_1>\pi_2=\pi_3$, the individual with the highest average payoff is certainly selected at birth. 
But selection at death will remove an individual carrying one of the two remaining strategies
with probability given by their abundance.
We find
\begin{subequations}
\label{eq:RPSLim02}
\begin{align}
&\,\lim_{\beta\to\infty}T_{2 \to 1} = \frac{i_{2}}{i_2+i_3},\\
\; \;
&\hbox{and}\nonumber\\
\; \;
&\,\lim_{\beta\to\infty}T_{3\to 1}  = \frac{i_{3}}{i_2+i_3}.
\end{align}
\end{subequations}
Obviously, we have $T_{2\to 1}+T_{3\to 1}=1$.
\item[(ii)] For $\pi_1=\pi_2>\pi_3$, the individual with the lowest payoff is selected for death with certainty, but selection at birth is still probabilistic.
It is easy to see that
\begin{subequations}
\label{eq:RPSLim03}
\begin{align}
&\,\lim_{\beta\to\infty}T_{3 \to 1} = \frac{i_{1}}{i_1+i_2},\\
\; \;
&\hbox{and}\nonumber\\
\; \;
&\,\lim_{\beta\to\infty}T_{3\to 2}  = \frac{i_{2}}{i_1+i_2}.
\end{align}
\end{subequations}
Thus, $T_{3\to 1}+T_{3\to 2}=1$. 
\item[(iii)] For $\pi_1=\pi_2=\pi_3$, selection is stochastic at birth and at death. In this case, we find
\begin{align}\label{eq:RPSLim04}
	\lim_{\beta\to\infty}T_{X \to Y} = \frac{i_X}{N} \frac{i_Y}{N}.
\end{align}
As expected, we have $\sum_{X,Y} T_{X \to Y} =1$. 
\end{itemize}
These results are valid if two or more payoffs are equal at a given lattice site, 
which is only obvious for $s=1$ and might not occur for any lattice site at all in the general case of $s\neq1$.

With \eq{RPS01}--\eq{RPS03}, we can compute the point sets in the simplex where two or all three average payoffs are equal, depending on the value of $s$. 
For $\pi_R=\pi_P=\pi_S$ this is only the center of the simplex, independent of $s$. 
For two payoffs being equal, we obtain the three linear equations 
\begin{align}
\nonumber
	\pi_R=\pi_S \quad & \hbox{at}\quad  i_P(s+2)  = {N-i_R(1-s)}{},\\
\label{eq:RPSPayEq}
	\pi_S=\pi_P \quad & \hbox{at}\quad  i_P(s-1)  = Ns-i_R(1+2s),\\
	\pi_R=\pi_P \quad & \hbox{at}\quad  i_P(2s+1)  =N(1+s)-i_R(2+s).\nonumber
\end{align}
However, it is not obvious at which of the discrete states $(i_R,i_P,i_S)$ we can observe equal average payoffs of two strategies if the losers payoff is not equal to $-1$, and especially if the system size is not a multiple of three, compare \fig{FIG4}. 


\begin{figure}[h]
\begin{center}
	\includegraphics[width=0.4\textwidth,angle=0]{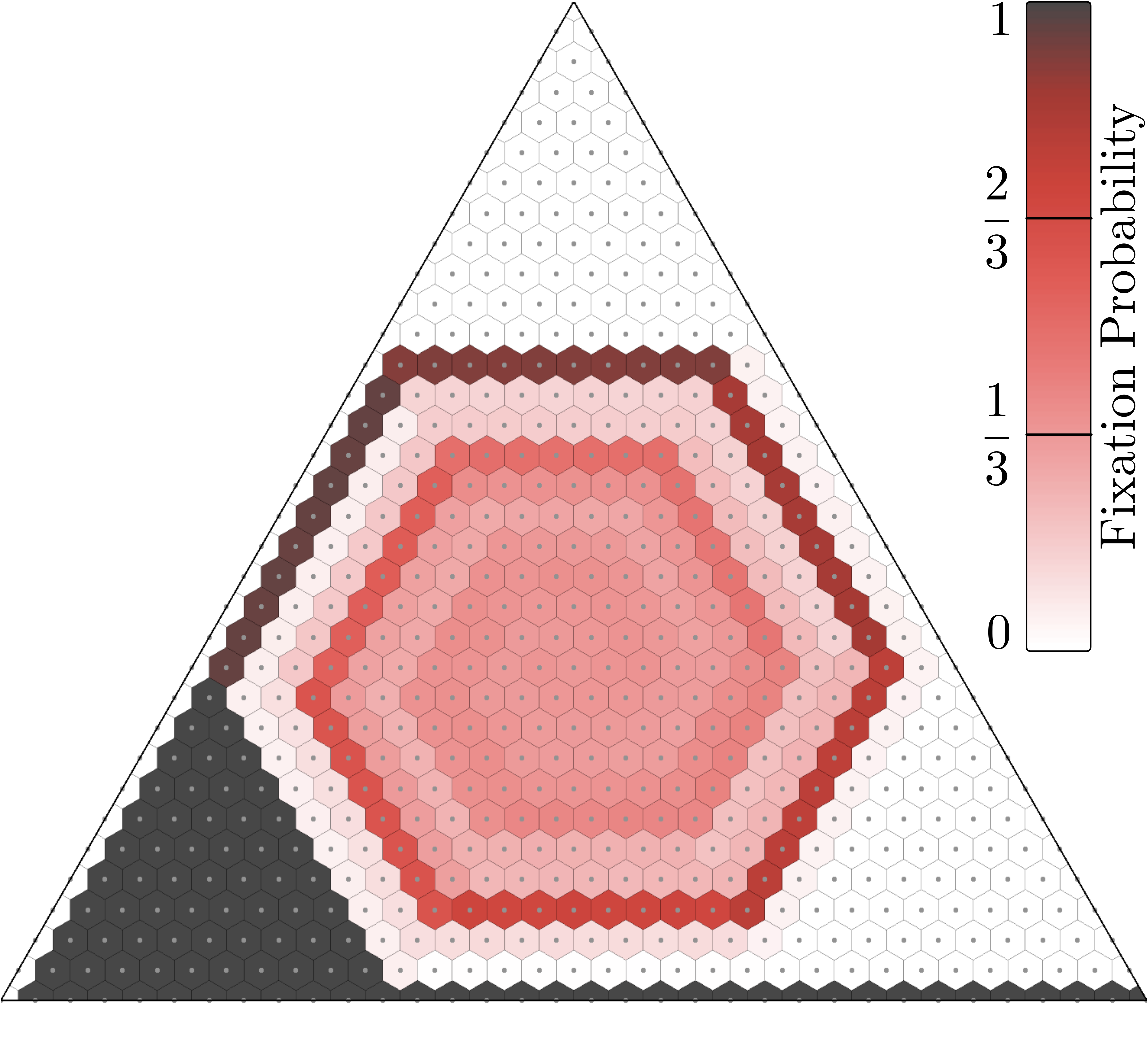}
\end{center}
\caption{ 
The probability that the system fixates at the lower left corner, \eg $\phi^R$, in a system with $N=33$, and $s=1$ for strong selection $\beta\to\infty$, depending on the initial state $(i_R, i_P, i_S)$. 
As discussed in the main text, fixation is stochastic.
Near the corners and along the edges fixation is deterministic. 
In a central area of the simplex, the system spirals out in a probabalistic fashion. 
The closer the initial condition is to the center, the closer the probability to get absorbed in a given state is to one third (fixation probabilities obtained from numerical simulations over $10^4$ realizations).  
}\label{fig:FIG5}
\end{figure}

For simplicity 
we thus concentrate on the case of $N$ being a multiple of three. 
In general, the strong selection behavior of the system is determined by the transition probabilities near the lines of equal average payoffs. 
When the population size $N$ is a multiple of three, the dynamics for the three different cases of $s=1$, $s<1$, or $s>1$, is as follows:

For $s=1$, we have 
$\pi_R=\pi_S$ at $i_P=N/3$, 
$\pi_S=\pi_P$ at $i_R=N/3$, and
$\pi_R=\pi_P$ at $i_S=N/3$. 
Hence, there is always stochastic movement induced by Eqs. \eq{RPSLim02} and \eq{RPSLim03} as well as \eq{RPSLim04}. 
Apart from these points, the direction of selection is indeed deterministic, which means that an individual with higher payoff always replaces an individual with a lower one. 
In certain regimes, near the corners of the simplex and along the edges the initial condition determines the final state where the system fixates. 
In a much larger area, however, the system fixates stochastically. 
In \fig{FIG5} we illustrate this by showing one fixation probability obtained from numerical simulations of the birth--death process.
Due to the symmetry of the system, this fixation probability can be either $\phi^R$, $\phi^P$, or $\phi^S$. 

\begin{figure}[h]
\begin{center}
	\includegraphics[width=0.4\textwidth,angle=0]{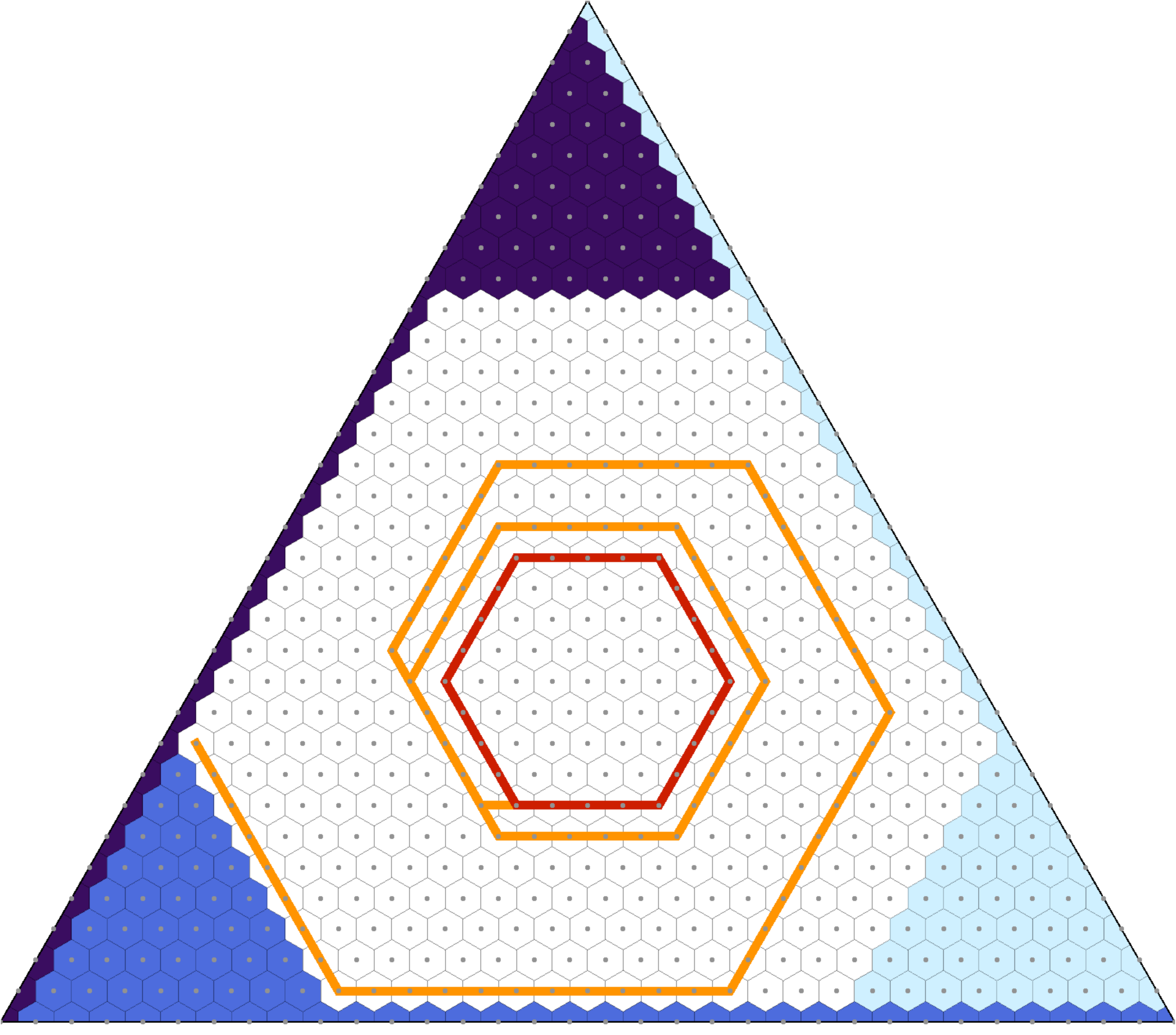}
	\end{center}
\caption{ 
For a population size of $N=33$ and $s=0.5$, the dynamics is such, that fixation does not occur in a central area, buy it is deterministic near the corners. 
For instance, any initial condition in the dark blue area will lead to the pure state at the lower left. 
The white area marks the sites from which, chosen as initial condition, the system does not fixate, but approaches a closed cycle. 
We show an example trajectory of the system that ends on the largest of these cycles (red hexagon) starting from the initial state $(i_R,i_P,i_S)=(1,9,23)$. 
Inside this largest cycle, there are other cycles such that every site except the center is already on such a cycle.  
}\label{fig:FIG6}
\end{figure}

For $s<1$, the lines where payoffs are equal rotate clockwise in our setup of cyclic dominance. 
In general, no states of the finite population system coincide with the lines of equal payoffs 
(except for special cases), see \fig{FIG4}. But as soon as the process crosses these lines, it changes direction. 
Near the corners and on the edges the system fixates deterministically, but in a central area the process spirals inwards if it does not hit the boundary of the system. 
However, it turns out that there is a largest limit cycle (LLC) depending on $s$ and $N$ and that there can be several other limit cycles (OLC) inside the LLC. 
We demonstrate this finding as it is obtained from numerical simulations in \fig{FIG6}, showing one sample trajectory that ends as the LLC. 
Note that the term cycle actually refers to a hexagon.

\begin{figure}[h]
\begin{center}
	\includegraphics[width=0.45\textwidth,angle=0]{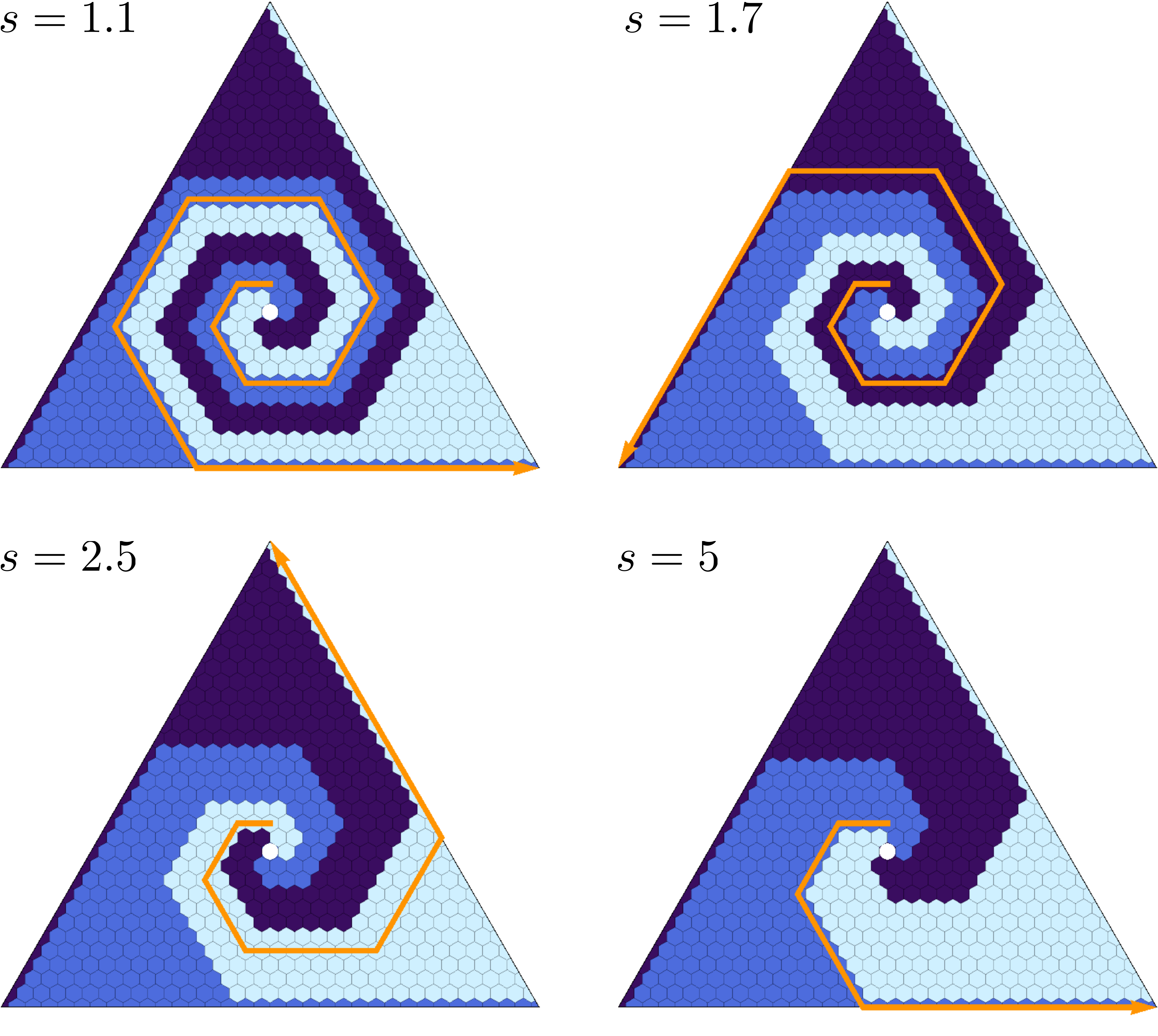}
	\end{center}
\caption{ 
Deterministic fixation for a population of size $N=33$ and different $s>1$. 
The dynamics is deterministic everywhere, except in the center (white dot), where all three payoffs are the same. 
Depending on the initial state, the system fixates to a given pure state. 
This is indicated by the three different colors at each state. 
For instance, when the system is initially at a site in dark blue, the system will fixate to the pure state at the lower left with probability one. 
We show one example trajectory (orange arrow) starting from the initial state $(i_R,i_P,i_S)=(10,13,10)$.
By increasing $s$, we can vary the state that is ultimately reached by the system. 
}\label{fig:FIG7}
\end{figure}

For $s>1$, as the lines of equal payoffs rotate counter clockwise (\fig{FIG4}) the system spirals outwards in a deterministic fashion. 
The movement is no longer deterministic only near the corners and along the edges but everywhere (except for the center), compare \fig{FIG7}. 
This means that if 
the modulus of 
the payoff for losing in our cyclic game, $s$, is larger than the payoff for winning, $1$, we observe strictly deterministic fixation depending on the initial condition. 

In this section, we have shown that a birth--death process with exponential payoff to fitness mapping and selection at birth and death is able to induce deterministic movement in the strong selection limit
even in $3 \times 3$ games.  
For the transition probabilities $T_{Y\to X}$ this limit can be performed analytically. 
It turns out that the microscopic dynamics is dependent only on the hierarchy of the average payoffs. 
As for finite systems an analytic description of the fixation probabilities (and times) is lacking, further examination of this system has to be numerical. 
Under strong selection the patterns that emerge show a very interesting regularity; it turns out that apart from system size, the results are dependent on the parameter $s$.

\section{Discussion}\label{sec:Disc}

The standard approaches to evolutionary game dynamics such as the Moran process or
pairwise comparison based on the Fermi rule lead to stochastic dynamics in finite populations \cite{traulsen:2005hp}. 
Even if the direction of selection becomes deterministic, the time scale typically remains stochastic and leads to a
distribution of the average fixation or mean exit times \cite{traulsen:2007cc}. 
Moreover, these standard approaches do not lead to a deterministic
direction of selection in games with more than two different strategies \cite{traulsen:2009aa}.
Here, we have introduced a process with selection at birth and at death. 
This process allows to interpolate between weak selection, usually considered in evolutionary 
biology, and arbitrary strong selection, such that in the extreme case the worst performing individual
is always replaced by a copy of the best performing individual. This kind of selection is sometimes
used in evolutionary optimization \cite{prugel:1994hb,mitchell:1996ef}. 
While the limiting case itself may not be of most interest for real biological or social systems, which
are always subject to stochastic noise, we discussed the most important
features of this limit. In particular, it reveals speed limits of evolutionary dynamics in $2 \times 2$ games that stochastic dynamics cannot cross and shows that
in games with more than two strategies, the limiting deterministic dynamics can have a crucial
dependence on the initial conditions. 

\section*{Acknowledgement}
	Financial support by the Emmy-Noether program of the DFG is gratefully acknowledged.

\end{document}